\documentclass{IEEEtran}

\usepackage{url} 
\usepackage{multirow}
\usepackage{fixltx2e}
\usepackage{enumerate}

  \usepackage[pdftex]{graphicx}
  \usepackage{epstopdf} 
  \graphicspath{{Figures/}}
  \DeclareGraphicsExtensions{.pdf,.jpeg,.png,.eps,.jpg}

\usepackage[cmex10]{amsmath}

\begin{document}

\title{Modular converter system for low-cost off-grid energy storage using second life Li-ion batteries}

\author{\IEEEauthorblockN{\thanks{Copyright \copyright~2014 IEEE}Christoph R. Birkl*, Damien F. Frost, Adrien M. Bizeray, Robert R. Richardson, and David A. Howey\dag \\}
\IEEEauthorblockA{Energy and Power Group, Department of Engineering Science,\\ University of Oxford, UK\\
Email: *christoph.birkl@eng.ox.ac.uk, \dag david.howey@eng.ox.ac.uk}}
%

\maketitle

\begin{abstract}
Lithium ion batteries are promising for small off-grid energy storage applications in developing countries because of their high energy density and long life. However, costs are prohibitive. Instead, we consider ``used'' Li-ion batteries for this application, finding experimentally that many discarded laptop cells, for example, still have good capacity and cycle life. In order to make safe and optimal use of such cells, we present a modular power management system using a separate power converter for every cell. This novel approach allows individual batteries to be used to their full capacity. The power converters operate in voltage droop control mode to provide easy charge balancing and implement a battery management system to estimate the capacity of each cell, as we demonstrate experimentally.
\end{abstract}

\IEEEpeerreviewmaketitle

\section{Introduction}\label{sec:Introduction}

The electrification of rural areas in developing countries ranks among the greatest humanitarian challenges of our time. Twenty percent of the world's population lack access to electricity due to deficiencies in infrastructure and financial means \cite{IEAWorldEnergyOutlook}. Energy storage technologies, particularly batteries, are key to providing independent electricity access where the grid is unavailable or weak, usually by means of solar photovoltaic (PV) systems. Currently, lead acid batteries are the most common technology for off-grid energy storage applications due to their low cost. However, lead acid batteries have low energy density (on the order of 40 W$\cdot$h/kg \cite{tarascon2001issues}), a short lifetime (100-800 cycles \cite{Ruetschi200433}) and high environmental impact if hazardous lead is released as a consequence of inadequate handling or disposal. The state of the art of secondary battery technology is lithium ion (Li-ion) with high energy density (approx. 130 W$\cdot$h/kg \cite{tarascon2001issues}) and long cycle life ($>$2000 cycles \cite{Wang20113942}). New Li-ion batteries are, however, not financially feasible for off-grid energy storage in developing countries due to their high cost. Meanwhile, the short life cycle of consumer electronics leads to the disposal of hundreds of tonnes of Li-ion batteries every year \cite{WasteBat}. Our analysis of 57 discarded Li-ion battery cells revealed that 50\%  of cells retained capacities of $>$70\%  of their nominal capacities and are, thus, still useable. This result is supported by tests conducted by Schneider et al. \cite{schneider2014classification}, who found that 45\%  of 227 analysed Li-ion cells were still useable. Ongoing long-term tests indicate that these cells remain operational for more than 160 deep cycles. 

\begin{figure}[!t]
	\centering
	\includegraphics[width=\linewidth]{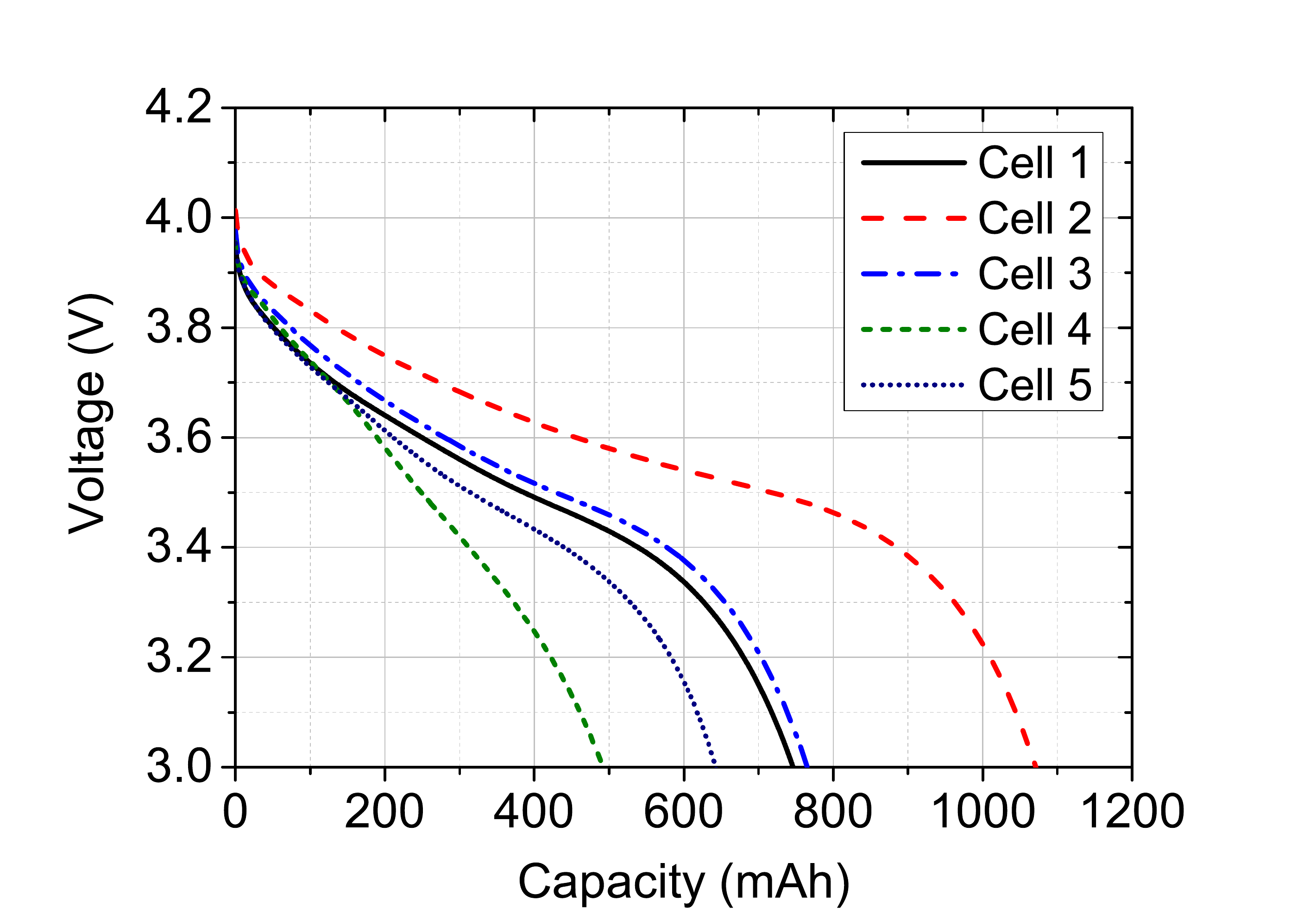}
	\caption{Discharge voltage vs capacity recorded during a constant current discharge at $I = 1250$ mA$\cdot$h for 5 cells retrieved from a single device.}
	\label{fig:CapComp}
\end{figure}

The challenge in second life applications for Li-ion batteries is identifying the useable cells and re-combining cells of various types and residual capacities into a functioning, safe device. Li-ion cells are generally connected in series, for higher battery voltage, or in parallel, for higher battery current. Large differences in cell capacity are problematic for series connections, since the same current passes through all cells and the total battery capacity is limited by the cell with the lowest capacity. Our analysis shows (see Section \ref{sec:Experimental}) that cell capacities can vary significantly even between cells recovered from the same device, as illustrated in Figure \ref{fig:CapComp}. In a series connection of the cells in Figure \ref{fig:CapComp}, the total battery capacity would be limited by Cell 4, which means that more than 50\% of the capacity of Cell 2 would be unutilized.

\begin{figure}[!t]
	\centering
	\includegraphics[width=\linewidth]{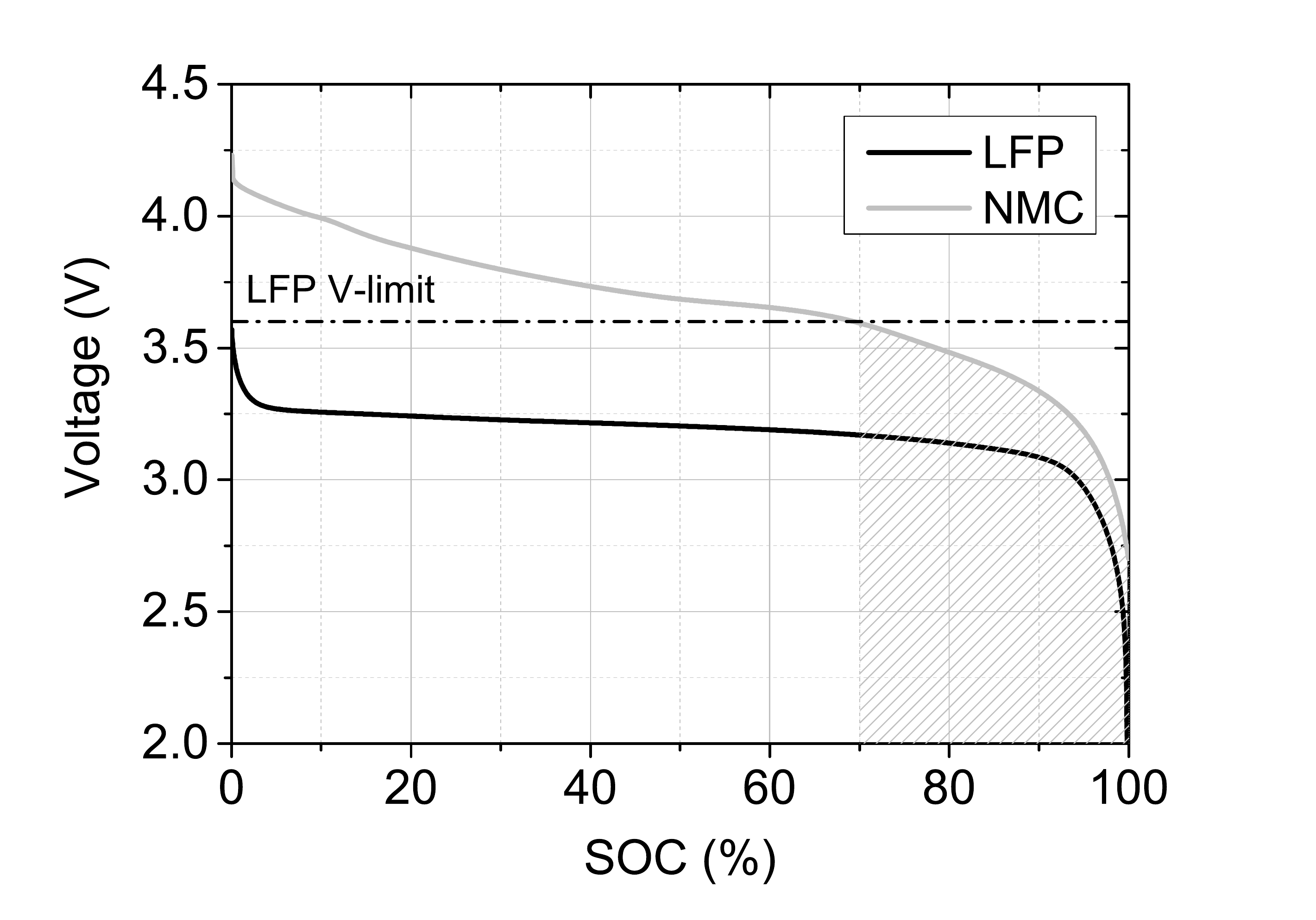}
	\caption{Voltage vs State of Charge (SOC) of LFP and NMC cells at 1C discharge rate. The shaded area indicates the capacity range of the NMC cell if constrained to the LFP voltage limit.}
	\label{fig:LFP-NMCcomp}
\end{figure}

For parallel connections of cells, differences in cell voltages are problematic, since all cells are tied to the same voltage and the total battery voltage is constrained by the cell with the lowest voltage limit. Cell voltages can vary significantly for different cathode chemistries. This is demonstrated in Figure \ref{fig:LFP-NMCcomp}, which shows the discharge curves of a LiFePO\textsubscript{4} (LFP) cell and a LiNiMnCoO\textsubscript{2} (NMC) cell under a current rate of 1C. A parallel connection of the two cells displayed in Figure \ref{fig:LFP-NMCcomp} would be constrained by the upper voltage limit of the LFP cell (3.6 V). This means that only approximately 30\% of the capacity of the NMC cell could be used. 

The above examples demonstrate the issues involved in constructing a battery pack from Li-ion cells of different types and capacities while utilizing each individual cell to its full potential. We address this challenge with a novel battery management system (BMS) comprised of multiple bi-directional dc-dc converters. These converters decouple the voltages and currents of the individual cells, and allow the ability to connect an arbitrary number of cells in a single device, thus scaling the total battery capacity as required. The BMS algorithms estimate the residual battery capacity of the connected cell. The estimated capacity is used to scale the amount of power each converter will provide. In this way, converters connected to large capacity cells will provide a greater share of the load current. As an additional benefit, this battery management technique minimizes the degradation of already worn cells by reducing the currents drawn from these cells \cite{choi2002factors}.

\section{Electrical design}\label{sec:ElectricalDesign}

\begin{figure}[!t]
	\centering
	\includegraphics[width=\linewidth]{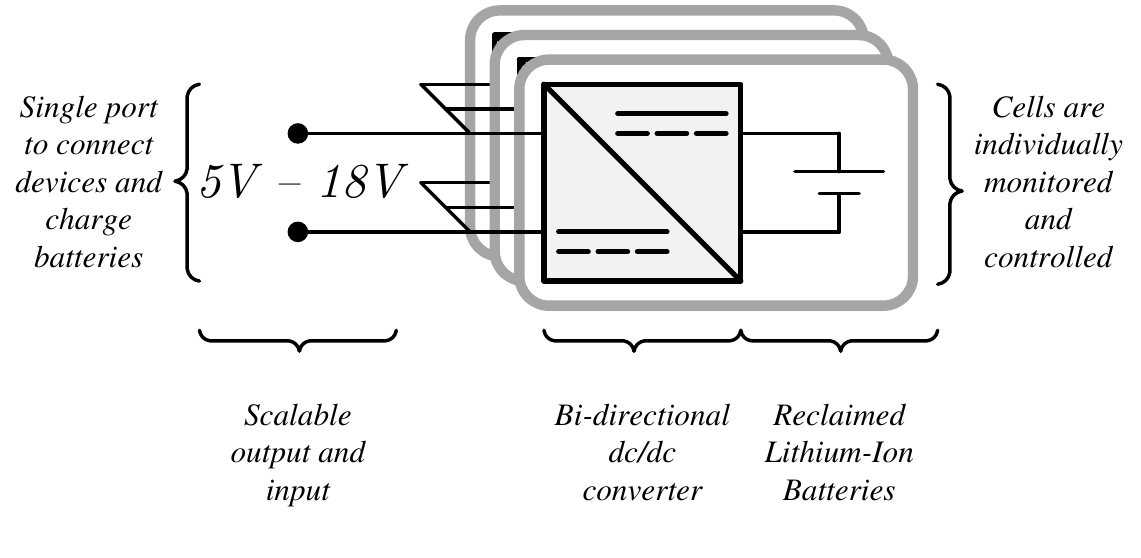}
	\caption{Schematic of the proposed system of using reclaimed Li-ion cells in a scalable energy storage system.}
	\label{fig:SystemSchematic}
\end{figure}

In order to maximize the remaining energy storage capacity in recovered Li-ion cells of varying degrees of degradation, the SOC of each cell must be monitored and controlled individually. We address this challenge by interfacing each cell with an individual power module. The proposed power module contains a small switch mode power supply (SMPS) which regulates the power in and out of the cell, a micro-controller which implements the control and BMS algorithms, and an output voltage bus that can be connected in parallel with other power modules to increase the energy storage capabilities of the entire system. Furthermore, the system is designed such that no communication between the power modules is necessary, however, load sharing is still achieved amongst the paralleled power modules. Figure \ref{fig:SystemSchematic} shows a schematic representation of the system.

\subsection{Module design}
\begin{figure}[!t]
	\centering
	\includegraphics[width=\linewidth]{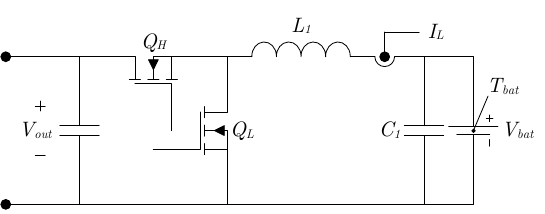}
	\caption{Circuit schematic of the power stage.}
	\label{fig:CircuitSchematic}
\end{figure}

Each power module contains a micro-controller which runs the BMS and controls the power flow in and out of the cell. As a proof of concept, a bi-directional half bridge dc-dc converter was used in the power stage \cite{fpe}. The converter measures the input and output voltages, as well as the inductor current and battery temperature. Figure \ref{fig:CircuitSchematic} shows a schematic of the power stage.

The output of each bi-directional half bridge dc-dc converter is a bi-directional power port which can be connected in parallel with other converters and be connected to a charging source. The charging source can be a grid connected power supply or a solar PV panel. In the case where the output is connected to a solar PV panel, the maximum power point (MPP) of the panel will change with temperature and solar irradiance conditions \cite{870713}. Therefore, the micro-controller will implement a perturb and observe maximum power point tracking (MPPT) algorithm \cite{653972} to track the maximum power of the panel.

Each converter acts independently of the others to share the load between cells in proportion to their capacity.

\subsection{Control}
\begin{figure}[!t]
	\centering
	\includegraphics[width=\linewidth]{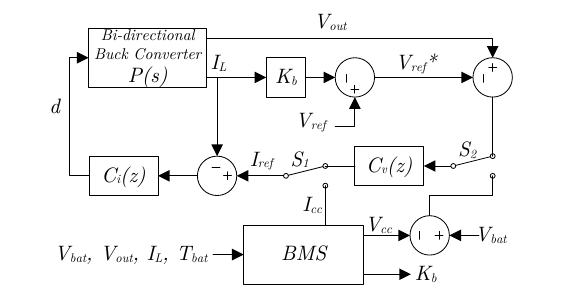}
	\caption{Simplified schematic of the control system within each power module.}
	\label{fig:ControlDiagram}
\end{figure}

Figure \ref{fig:ControlDiagram} shows a diagram of the controllers implemented in the micro-controller. There are three main operating modes of the converter. The discharge mode, Mode 1, provides a nominal $12$ V to the output of the converter. The charging modes, Modes 2 and 3, are activated once the output of the converter is connected to a voltage source between 14 V and 20 V.

\subsubsection{Mode 1: Discharge}

In Mode 1, the Li-ion cell is being discharged into a load connected to the output terminals of the power module. In this mode, the control flow switches $S_1$ and $S_2$ are in the “up” position, as shown in Figure \ref{fig:ControlDiagram}. The inner current control loop with controller $C_i(z)$ and the outer voltage control loop with controller $C_v(z)$ work together to maintain a voltage $V_{ref}*$ at the output terminals.

Current sharing of the load is achieved using voltage droop control  \cite{perkinson1995current}. As shown in Figure \ref{fig:ControlDiagram}, a nominal voltage reference, $V_{ref}$, is modified proportionally to the output current of the converter. The proportionality constant, $K_b$, by which the output voltage reference is modified is determined by the BMS. $K_b$ is inversely proportional to the battery capacity. Thus, power modules which have larger Li-ion cells will provide more current to the load than the modules with smaller capacity Li-ion cells.

While the converter is operating, the BMS monitors the battery, ensuring that it is operating within its safety limits. The BMS also performs a simple capacity estimation to determine the parameter $K_b$.

\subsubsection{Mode 2: Charging with Constant Current}

Mode 2 is activated when the power module’s output is connected to a 14 V and 20 V charging source. In this mode, the control flow switch $S_1$ is in the “down” position, and the voltage controller, $C_v(z)$, is off.

The current reference for the converter is provided by the BMS which is implementing a perturb and observe MPPT algorithm \cite{653972}. The current reference will be proportional to the capacity of the cell, and will vary according to the MPPT algorithm. If a new cell is attached, the current reference will be set to its minimum value. In the case where the converter is connected to a grid-connected voltage source, the MPPT algorithm will request the maximum charging current for the cell that is being charged. 

During Mode 2, the BMS monitors the battery voltage and switches to Mode 3 when the upper voltage limit of the battery is reached.

\subsubsection{Mode 3: Charging with Constant Voltage}

In Mode 3, the control flow switch $S_1$ is in the “up” position, and the control flow switch $S_2$ is in the “down” position. The BMS provides a voltage reference which is compared to the battery voltage. The voltage controller $C_v(z)$ now controls the battery voltage, instead of the output voltage.

The BMS will determine when the battery is fully charged by monitoring $I_L$ and comparing it to a cut-off current. It will also determine if there is enough power from a charging source such as a solar PV panel by ensuring that $V_{out}$ remains above 14 V, while $I_L$ is still charging the battery.

\section{Software and algorithms}\label{sec:Software}

The algorithms designed for this BMS serve two main purposes:
\begin{enumerate}[A)]
	\item Condition monitoring for safe operation 
	\item Current control and balancing of individual cells
\end{enumerate}

\subsection{Condition monitoring for safe operation}

Each cell is equipped with a temperature, voltage and current sensor. Upper and lower safety limits on those parameters are given in Table \ref{tab:SafetyLimits}. Safety limits are based on a review of manufacturer specifications of Li-ion cells commonly used in electronic devices. Temperature limits are similar for most cell types. The lower temperature limit is more conservative for charging, since very low temperatures can trigger the formation of lithium plating and dendrite growth, which can lead to internal short circuits. Voltage limits depend on the cathode chemistry. LFP cells have a generally lower range of operating voltage than most other chemistries (3.6 V to 2.0 V). These cells are identified during the initial characterisation, by detecting the sharp voltage gradient during charge, when approaching their maximum voltage of 3.6 V. For all other chemistries, the most conservative voltage range of 4.2 V to 3.0 V is applied. The current is limited to 3.0 A, which is well within the operating range of 2500 mA$\cdot$h to 2900 mA$\cdot$h cells.

\begin{table}[!h]
	\begin{center}
		\caption{Safety limits.}\label{tab:SafetyLimits}
		\begin{tabular}{ccc}
			\hline
			Parameter & Upper Limit & Lower Limit\\
			\hline\hline
			\multirow{2}{*}{Temperature:} & \multirow{2}{*}{60$\,^{\circ}\mathrm{C}$} & charge: 0$\,^{\circ}\mathrm{C}$\\
			& & discharge: $-10\,^{\circ}\mathrm{C}$\\
			Voltage: & 3.6 V - 4.2 V & 2.0 V - 3.0 V\\
			Current: & 3.0 A & --\\
			\hline
		\end{tabular}
	\end{center}
\end{table}

These safety limits are continuously monitored, by sampling at a frequency of $5$ kHz. Breaching any safety limits triggers an immediate shut down of the power module, isolating the affected cell. 

\subsection{Algorithms for current control and cell balancing}

As described in Section \ref{sec:ElectricalDesign}, the bi-directional dc-dc converters allow independent current control on each Li-ion cell. In order to optimally utilize their capacities, the current through each cell must be controlled such that all cells discharge simultaneously. This means that a given load current must be provided by individual cells according to their capacities; i.e. higher capacity cells must be subjected to higher currents than lower capacity cells. However, cell capacities are not known for reused cells of different types and states of health. We address this problem with a novel algorithm that estimates battery capacities by means of a comparative/iterative Coulomb counting approach \cite{Ng20091506}. The capacity of a cell at a given discharge current can be calculated according to:

\begin{equation}\label{eq:coulombCounting_Continuous}
	Q = \int_{t=0}^t I\left(t\right) dt
\end{equation}

where I is current and t discharge time. For discrete time intervals $k$, Equation \ref{eq:coulombCounting_Continuous} can be expressed as 

\begin{equation}\label{eq:coulombCounting_Discrete}
	Q = \sum_{k=1}^{N} I_k \Delta t
\end{equation}

The cell capacity can thus be calculated from accurate current measurements performed at small time intervals. We employ this capacity measurement along with the average current during discharge to determine the parameter $K_b$, used in the voltage droop controller as shown in Figure \ref{fig:ControlDiagram}.

\begin{figure}[!t]
	\centering
	\includegraphics[width=70mm]{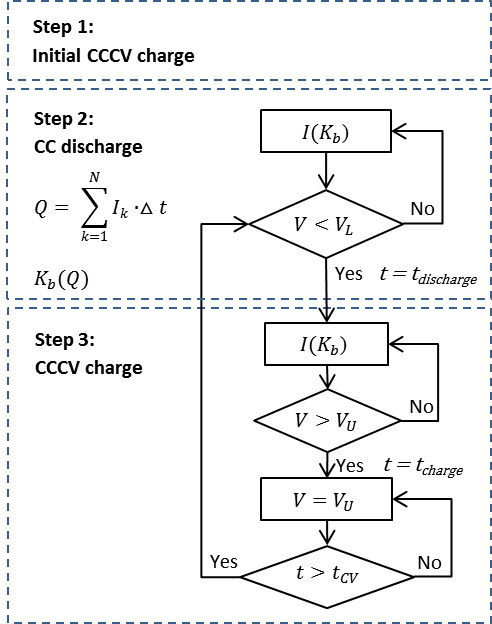}
	\caption{BMS algorithm for conditioning cycle.}
	\label{fig:BMSAlgo}
\end{figure}

Capacity measurements and current scaling are implemented for all cells in the battery and the computations are repeated with every charge and discharge cycle, as illustrated in the flowchart shown in Figure \ref{fig:BMSAlgo}. The algorithm is initiated with a first constant current constant voltage (CCCV) charge to balance the cells at a uniform state of charge (SOC) (Step 1 in Figure 6). All cells are charged with equal currents to their maximum voltages, which are held until a predefined time limit is exceeded. After that, the cells are discharged with equal currents, until the cut-off voltages are reached (Step 2 in Figure \ref{fig:BMSAlgo}). Measuring the time of this first discharge cycle allows calculating the cell capacities and provides an initial estimate of $K_b$.

For the first CCCV charge cycle, the $K_b$ value calculated in Step 2 is used to correct the charge current. The CCCV charge is conducted as described in Step 3 of Figure \ref{fig:BMSAlgo}. The charge capacity of cycle 1 is calculated by coulomb counting, in the same manner as the discharge capacity. $K_b$ is updated at the end of the charge cycle.

Upon start-up of the device, a full charge-discharge-charge cycle is conducted (Steps 1 to 3, Figure \ref{fig:BMSAlgo}) in order to adjust cell currents and synchronize charge and discharge times. During ordinary operation the current correction factors are updated for each cell by continuous comparison of the charged/discharged energy with that of the previous charge/discharge cycle. In this manner, $K_b$ continually reflects the changing capacity as the cell degrades over time.

\begin{figure}[!t]
	\centering
	\includegraphics[width=80mm]{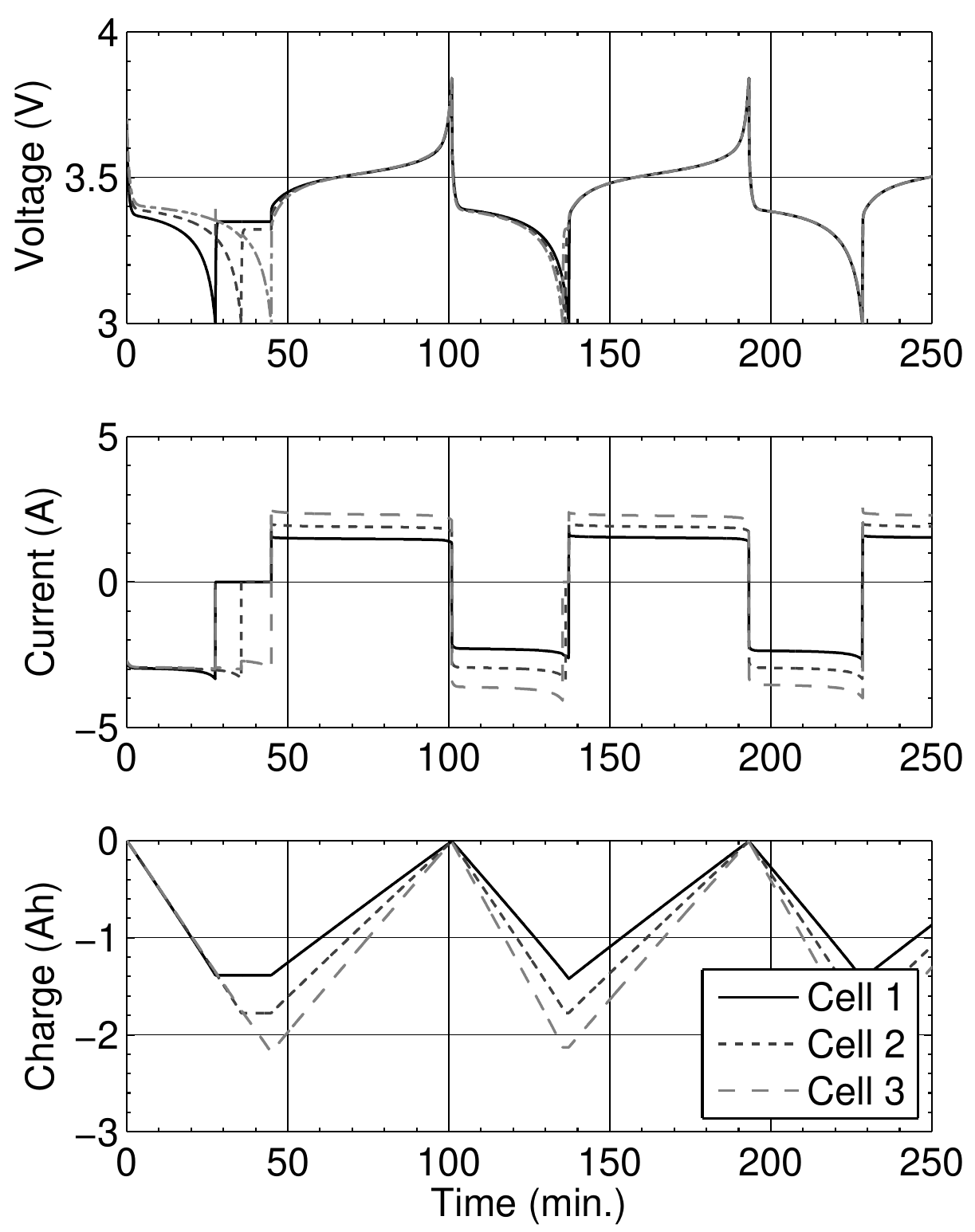}
	\caption{Simulation of the BMS algorithm synchronizing three Li-ion cells of different capacities; Cell 1: 1600 mA$\cdot$h, Cell 2: 2000 mA$\cdot$h and Cell 3: 2400 mA$\cdot$h.}
	\label{fig:BMSSim}
\end{figure}

The above described algorithm was implemented in MATLAB Simulink. Figure \ref{fig:BMSSim} demonstrates how the algorithm synchronises discharge and charge cycles of three cells with different capacities by adjusting the current load on each cell in proportion to their capacity. The cycle times of the three cells converge after the first two discharge-charge cycles. Cell capacities used for the simulation are 1600 mA$\cdot$h for Cell 1, 2000 mA$\cdot$h for Cell 2 and 2400 mA$\cdot$h for Cell 3. The standard Li-ion battery model of Simscape SimPowerSystems is used to emulate the battery voltage in response to a current load. For the first cycle of the simulation, a discharge power of 30 W and a charge power of 20 W are divided equally among the three cells. The cell capacities are estimated during each successive charge and discharge cycle to vary the parameter $K_b$. The simulation validates the feasibility of the BMS algorithm for synchronizing cells of different capacities by active current control. Further simulations are required in order to validate the long-term stability of the algorithm and its capability to synchronize cells of different chemistries. 

\section{Experimental}\label{sec:Experimental}

\subsection{Characterisation of recovered Li-ion cells}\label{sec:Characterisation}

The viability of using Li-ion battery cells recovered from discarded electronic devices in a second life application was investigated in a series of experiments on a total of 57 Li-ion cells. The tests were performed with an 8 channel MGP-205 battery tester by BioLogic and an 8 channel Battery Test System by Neware. Tested cell chemistries included LiNiMnCoO\textsubscript{2} (NMC), LiCoO\textsubscript{2} (LCO) and LiNiCoAlO\textsubscript{2} (NCA) in both cylindrical 18650 format and pouch format. Nominal capacities of the tested cells were in the range of 2500 mA$\cdot$h to 2900 mA$\cdot$h.

The test procedure consisted of the following steps:

\begin{enumerate}
	\item Visual inspection
	\item Voltage measurement
	\item Initial charge/discharge cycle:
	\begin{enumerate}
		\item Constant current charge, rate: $C/2$
		\item Constant voltage charge for $t = 20$ min
		\item Constant current discharge, rate: $C/2$
	\end{enumerate}
	\item Initial Capacity test:
	\begin{enumerate}
		\item  Constant current charge, rate: $C/2$
		\item  Constant voltage charge, current limit: $50$ mA
		\item  Constant current discharge, rate: $C/2$
	\end{enumerate}
	\item  Cycling and capacity tests
	\begin{enumerate}
		\item Charge and discharge cycles in sets of 20 cycles, as described in step 3.
		\item Capacity tests every 20 cycles, as described in step 4. 
	\end{enumerate}
\end{enumerate}

\begin{figure}[!t]
	\centering
	\includegraphics[width=\linewidth]{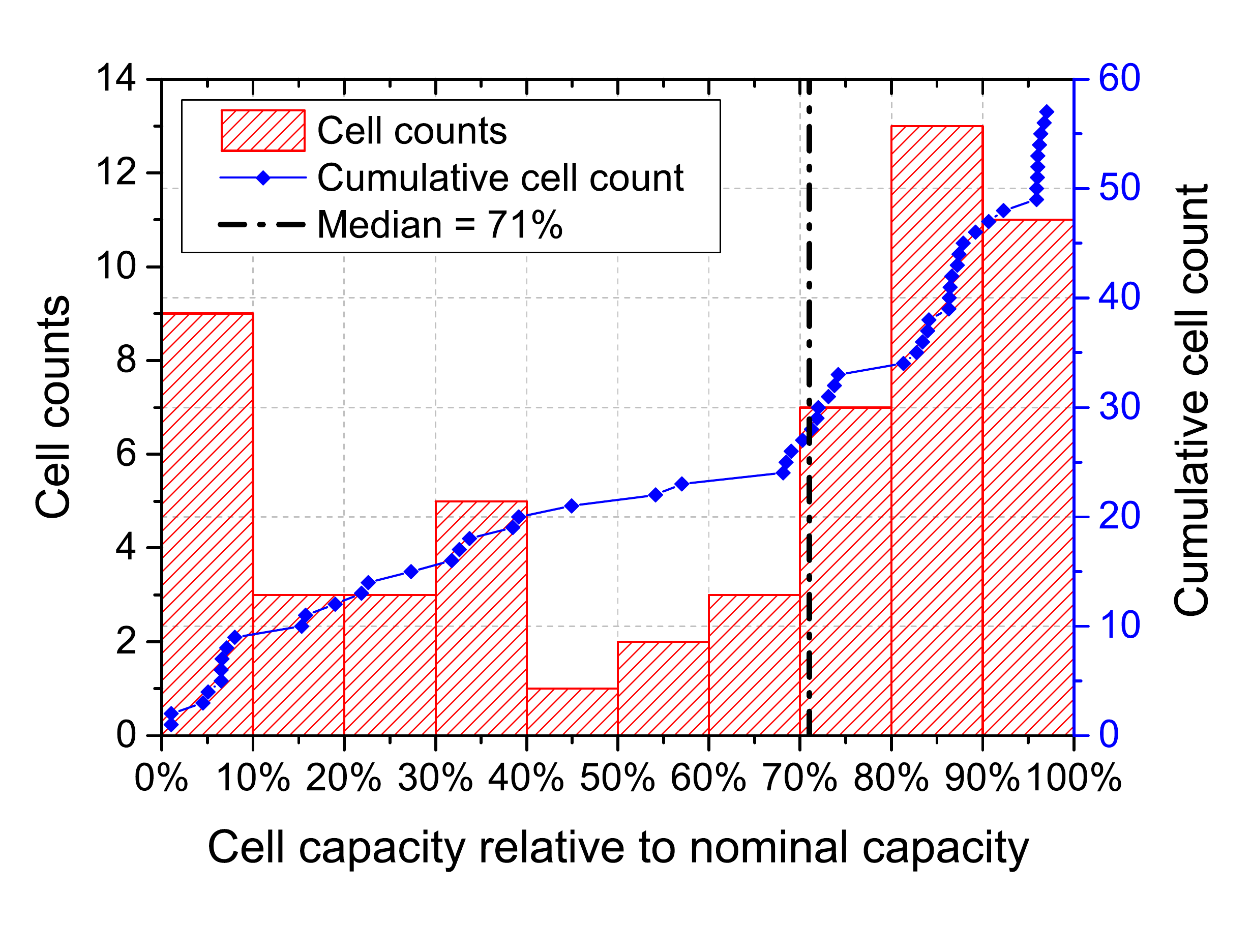}
	\caption{Analysis of remaining useful cell capacity.}
	\label{fig:CellStats}
\end{figure}

No superficial signs of physical defects were found on any of the 57 cells during the visual inspection. Initial voltage measurements showed that 2 cells were at voltages $<$0.7 V  and thus considered defective. Initial charge and discharge cycles and capacity tests were performed on the remaining 55 cells. Operating limits and nominal cell capacities were obtained from data sheets. Current rates for charge and discharge of the individual cells were calculated from their respective nominal capacities, which ranged from 2500 mA$\cdot$h to 2900 mA$\cdot$h. The results of the initial capacity tests are summarized in Figure \ref{fig:CellStats}. It was found that half of the tested cells retained capacities of $>$70\% of their nominal capacities and 63\%  of cells retained capacities of $\geq$ 50\%  of their nominal capacities.

\subsection{Hardware tests}
Three circuits described in Section \ref{sec:ElectricalDesign} were tested with their outputs in parallel. For safety reasons, Li-ion battery behaviour was simulated by a BioLogic MGP-205 battery tester. One channel was connected to each power module and the BioLogic MGP-205 was operated as a voltage source. The voltage profiles used to simulate Li-ion battery cells were recorded on real cells. The load was provided by a 47 $\Omega$ resistor, connected to the paralleled outputs of the power modules. The output voltage of the three modules was measured with a BioLogic SP150 potentiostat and VMP3B 10 A booster. 

\begin{figure}[!t]
	\centering
	\includegraphics[width=60mm]{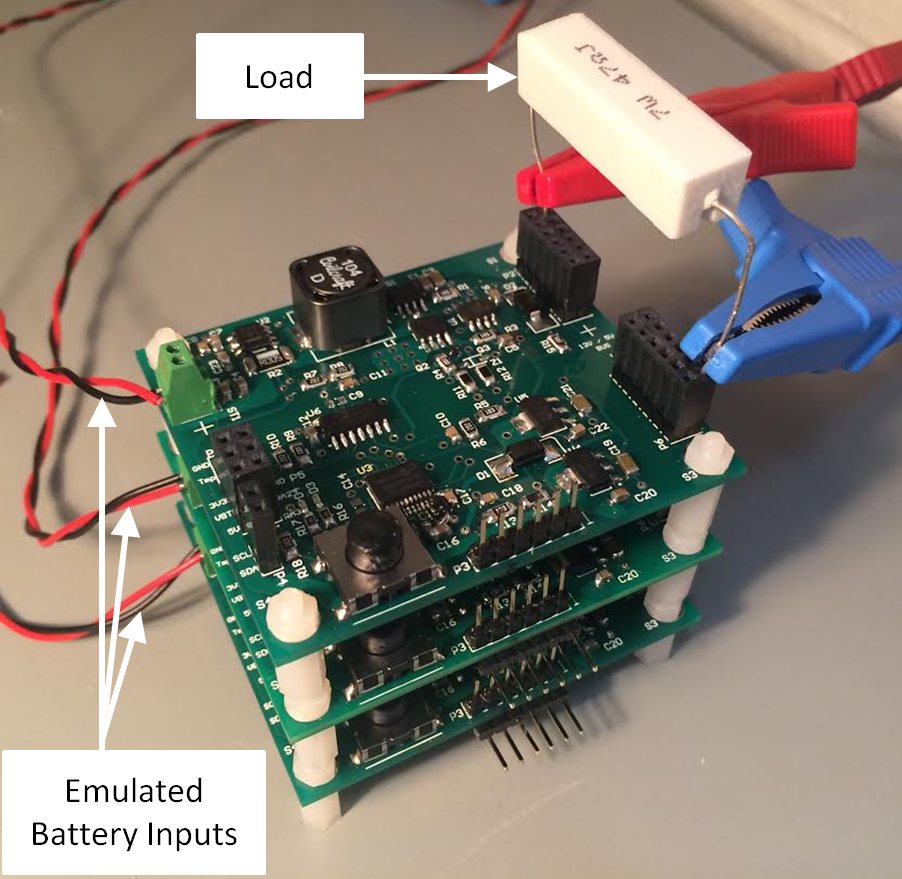}
	\caption{Experimental setup to test three paralleled power modules.}
	\label{fig:TestSetup}
\end{figure}

The objective of the hardware tests was to evaluate three crucial functions of the power modules, namely their capability to

\begin{enumerate} 
	\item operate Li-ion cells of different capacities under different currents
	\item operate Li-ion cells of different chemistries, i.e. different cell voltages and respective operating limits
	\item compensate for the failure of individual cells, while providing a stable output voltage of 12 V $\pm 1$.
\end{enumerate}
Table \ref{tab:HWTestOverview} gives an overview of the three experiments conducted for the validation of hardware functionalities.

\begin{table}[!h]
	\begin{center}
		\caption{Hardware test overview.}\label{tab:HWTestOverview}
		\begin{tabular}{cccc}
			\hline
			Test & Objective & {\parbox{1.5cm}{\centering Simulated chemistry}} & {\parbox{1.5cm}{\centering Simulated cell capacity}}\\
			\hline\hline
			\multirow{3}{*}{I} & \multirow{3}{*}{\parbox{2.4cm}{\centering Compensation of capacity difference by current control}} & NMC & 75 mA$\cdot$h \\
			& &	NMC	& 100 mA$\cdot$h\\
			& &	NMC & 150 mA$\cdot$h\\
			\hline
			\multirow{3}{*}{II} & \multirow{3}{*}{\parbox{2.4cm}{\centering Compensation of voltage difference by current control}} & NMC & 220 mA$\cdot$h \\
			& &	LFP	& 220 mA$\cdot$h\\
			& &	NMC & 220 mA$\cdot$h\\
			\hline
			\multirow{3}{*}{III} & \multirow{3}{*}{\parbox{2.4cm}{\centering Compensation of cell failure}} & NMC & 110 mA$\cdot$h \\
			& &	LFP	& 275 mA$\cdot$h\\
			& &	NMC & 275 mA$\cdot$h\\
			\hline
		\end{tabular}
	\end{center}
\end{table}

For Test I, the MPG-205 was used to simulate a voltage profile previously recorded on a Li-ion cell. Three voltage profiles were created and emulated on channels 1-3: 75 mA$\cdot$h, 100 mA$\cdot$h and 150 mA$\cdot$h, respectively. These different emulated capacities represent the differences in nominal capacity and/or state of health (SOH), i.e. capacity fade, of the reclaimed Li-ion cells. Discharge currents on individual cells are regulated by the BMS and determined by the estimated cell capacity, as described in Section \ref{sec:Software}. In reality, the voltage profile of a Li-ion cell changes as the cell degrades. Therefore, voltage profiles should ideally be recorded at the corresponding SOH. However, the small differences in curvature of the voltage profile as a result of degradation were neglected for this experiment and the same voltage profile was used for all channels. The objective of Test I was to demonstrate the synchronised discharge of three cells of different capacities under constant output voltage, enabled by appropriate adjustment of discharge currents performed by the parallel configuration of three converter modules.  

For Test II, the MPG-205 was used to simulate the voltage profiles previously recorded on an NMC cell, an LFP cell and an NCA cell. The capacity of all cells emulated on channels 1-3 was normalized to 220 mA$\cdot$h. The objective of Test II was to demonstrate the synchronised discharge of three Li-ion cells of different chemistries and safety limitations under constant output voltage, achieved by appropriate adjustment of discharge current. 

For Test III, the MPG-205 was used to emulate three cells of different chemistries (equivalent to Test II) with one discharge profile (of the NMC cell) significantly shorter than the others, simulating a failing cell. In the case of a failing cell, the remaining operational cells have to take over the load of the failed cell in order to ensure continued operation of the device. The objective of Test III was to prove the capability of the hardware to compensate for failing cells by appropriate adjustment of discharge currents of the remaining cells, while maintaining a constant output voltage.

\section{Test Results}\label{sec:TestResults}

\subsection{Test I: Compensation for different cell capacities}

\begin{figure}[!t]
	\centering
	\includegraphics[width=\linewidth]{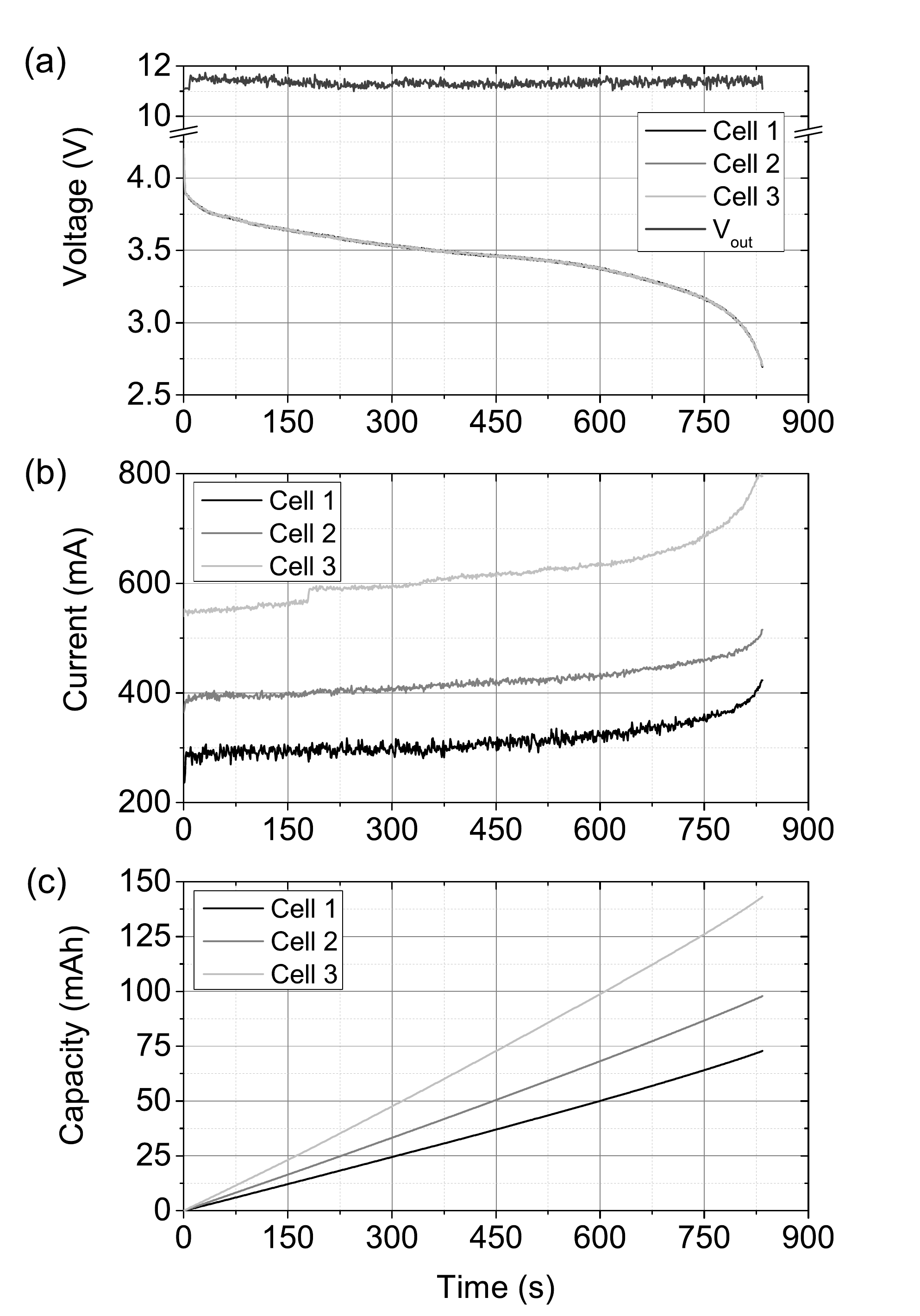}
	\caption{Test I - compensation for different cell capacities. (a) cell and output voltages (b) cell currents and (c) discharge capacities. All cells represent NMC cell chemistry.}
	\label{fig:DiffCapTest_VIQ_I}
\end{figure}

The results of Test I are illustrated in Figure \ref{fig:DiffCapTest_VIQ_I}(a), (b) and (c). Figure \ref{fig:DiffCapTest_VIQ_I}(a) clearly shows that all cells are following the same voltage profile (recorded on an NMC cell under a discharge rate of 4C). For these experiments, the cell capacity was provided to the BMS in advance. The different capacities of the three cells are illustrated in Figure \ref{fig:DiffCapTest_VIQ_I}(c); namely 75 mA$\cdot$h for Cell 1, $100$ mA$\cdot$h for Cell 2 and $150$ mA$\cdot$h for Cell 3. In order to synchronize the discharge of three cells of different capacities connected in parallel, the individual cell currents must be controlled appropriately. This is displayed in \ref{fig:DiffCapTest_VIQ_I}(b). The three paralleled converter modules actively regulated the discharge currents in proportion to the cell capacities and achieved a synchronized discharge, while maintaining a constant output voltage, as shown in  Figure \ref{fig:DiffCapTest_VIQ_I}(a), with an average value of $11.35$ V and a variance of $17$ mV.

\subsection{Test II: Compensation for different cell voltages (chemistries)}

\begin{figure}[!t]
	\centering
	\includegraphics[width=\linewidth]{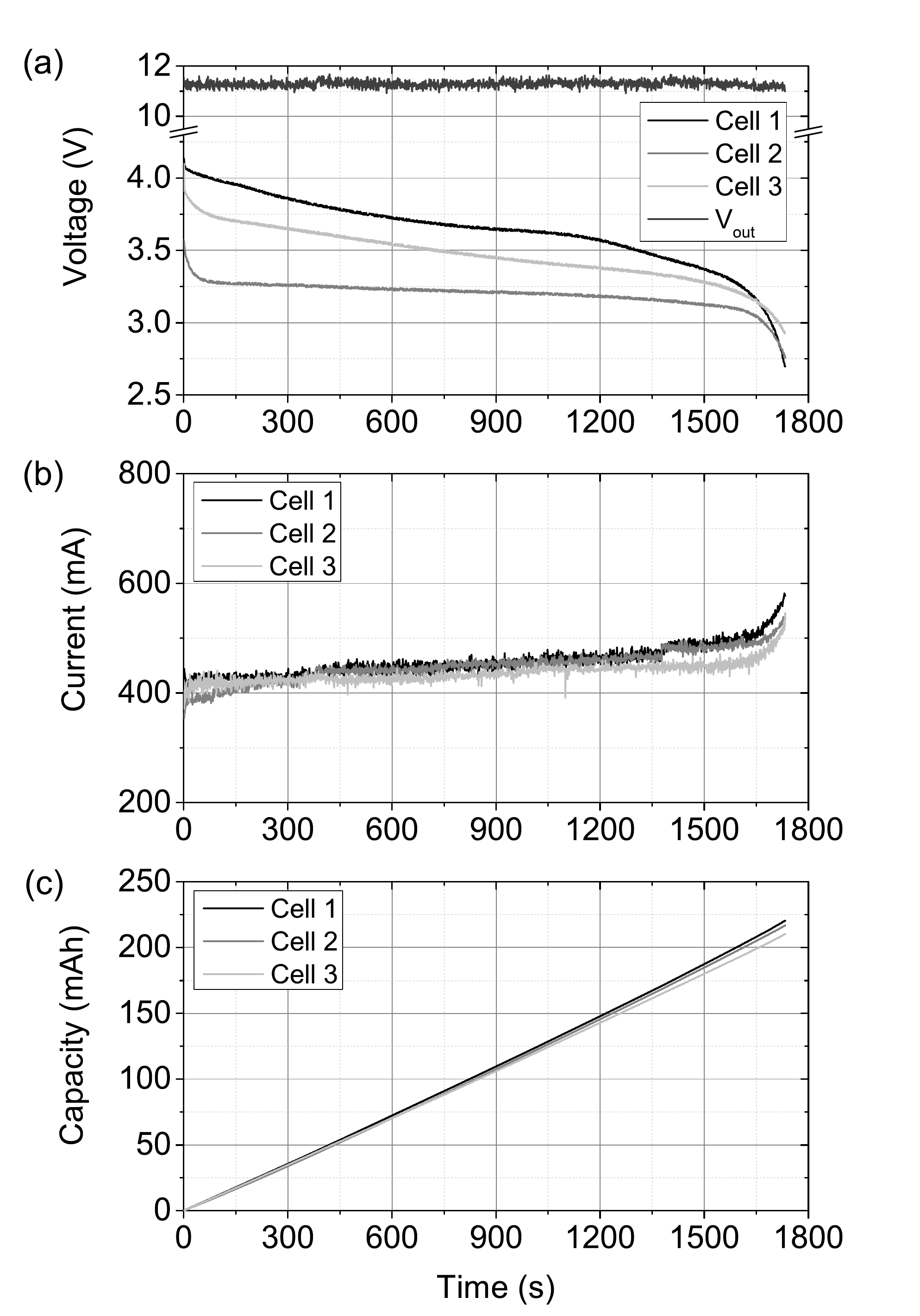}
	\caption{Test II - compensation for different cell voltages. (a) cell and output voltages (b) cell currents and (c) discharge capacities. Cell 1 represents NMC, Cell 2 LFP and Cell 3 NCA cell chemistries.}
	\label{fig:DiffChemTest_VIQ_II}
\end{figure}

Figures \ref{fig:DiffChemTest_VIQ_II}(a), (b) and (c) show the results of Test II. The simulated cell chemistries are NMC, LFP and NCA (voltage profiles recorded under 1C discharge). All cells are simulated with a capacity of $220$ mA in order to emphasize the effects of different cell voltages, which are illustrated in Figure \ref{fig:DiffChemTest_VIQ_II}(a). The power modules allow the different cell types (i.e. different cell voltages) to be connected in parallel and operated within their safe voltage limits. Figure \ref{fig:DiffChemTest_VIQ_II}(b) and (c) illustrate how the three cells are discharged simultaneously with equal currents. The output voltage was maintained at an average value of $11.29$ V and a variance of $17$ mV. 

\subsection{Test III: Compensation for cell failure}

\begin{figure}[!t]
	\centering
	\includegraphics[width=\linewidth]{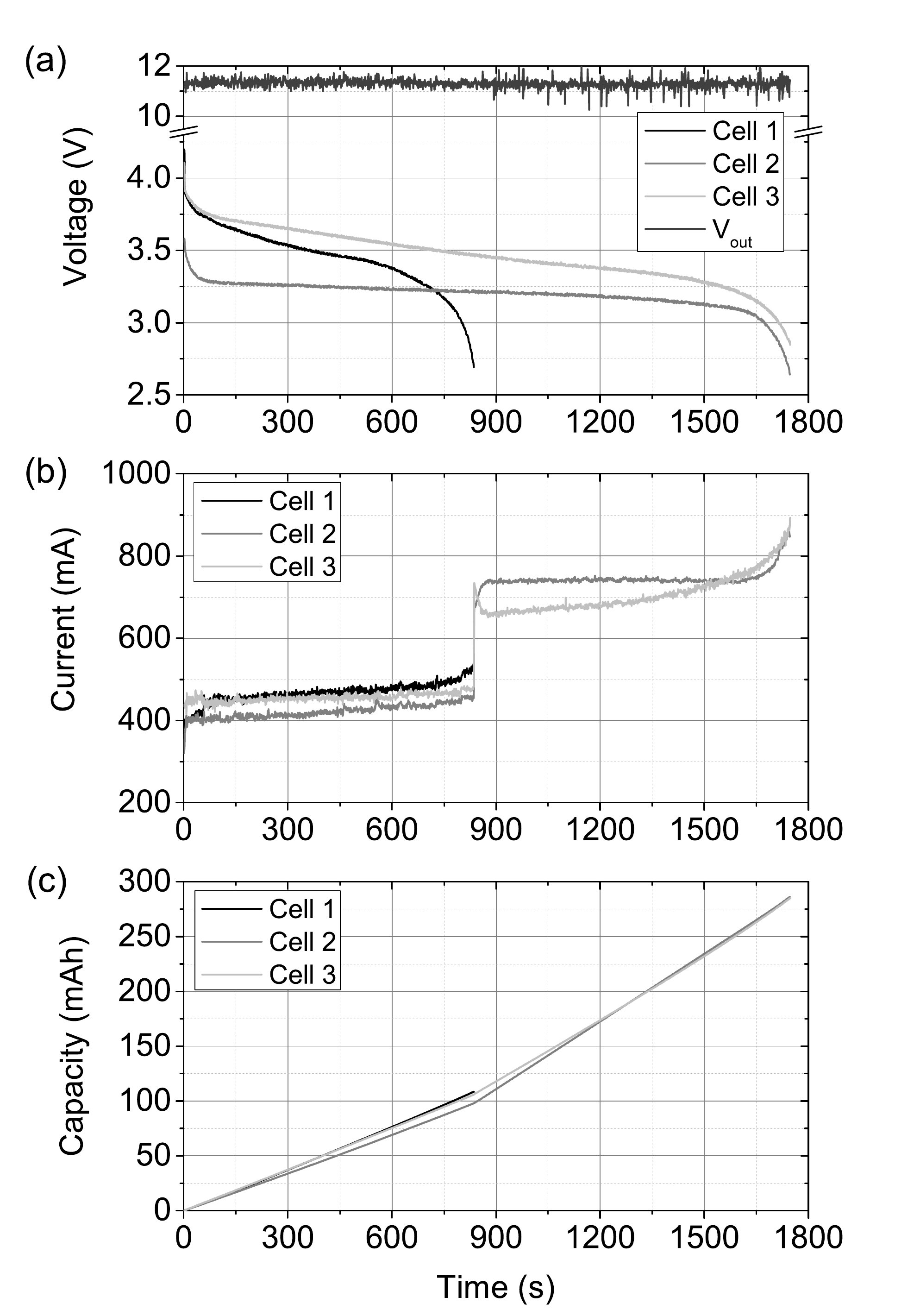}
	\caption{Test III - compensation for cell failure. (a) cell and output voltages (b) cell currents and (c) discharge capacities. Cell 1 represents NMC, Cell 2 LFP and Cell 3 NCA cell chemistries.}
	\label{fig:FailureTest_VIQ_III}
\end{figure}

The results of Test III are summarised in Figure \ref{fig:FailureTest_VIQ_III}(a), (b) and (c). The same cell types as in Test II were simulated. However, the voltage profile of the NMC cell was recorded under a discharge rate of $2 C$ (as opposed to the $1 C$ discharge rate on the LFP and NCA cells) and the resulting difference in discharge time was not compensated by current control so as to simulate the failure of a cell. Figure \ref{fig:DiffChemTest_VIQ_II}(a) demonstrates the different discharge periods (roughly half the time for the NMC cell) by means of the cell voltages. Figure \ref{fig:FailureTest_VIQ_III}(b) illustrates how the failing of the NMC cell (current goes to zero) was compensated by an increase in the LFP and NCA cell currents. This rapid change of current load on the LFP and NCA cells did not affect the output voltage to any significant extent, as apparent in Figure \ref{fig:FailureTest_VIQ_III}(a). The output voltage during Test III remained at an average value of $11.31$ V with a variance of $34$ mV. Figure \ref{fig:FailureTest_VIQ_III}(c) demonstrates how the LFP and NCA cells were discharged by the same amount despite the disruption caused by the failure of the NMC cell.

\section{Financial Feasibility Assessment}\label{sec:Finance}

\subsection{System cost estimate}

The greatest financial advantage of the proposed system over other battery electric energy storage devices lies in the low cost of second-life Li-ion battery cells. The content of profitably recyclable raw materials in novel Li-ion  cell types is low and discarded cells therefore hold little or no value \cite{wang2014economic}. However, environmental risks associated with the disposal of large quantities of batteries provide a socio-political incentive to enforce the recycling of Li-ion cells. Western market economies are therefore starting to hold battery manufacturers and retailers responsible for the life cycle costs invoked by their products. For example, the UK Environmental Protection Act 2009 No. 890 obligates battery manufacturers and retailers to finance the net cost arising from the collection, treatment and recycling of waste batteries. This provides a great incentive and opportunity to prolong the useful lifetime of rechargeable batteries at virtually no additional cost, since the collection and subsequent recycling of the batteries is already financed by manufacturers and retailers. For this reason, we assume in our financial feasibility assessment that the second-life Li-ion batteries do not add cost to the proposed energy storage device. 

\begin{table}[!h]
	\begin{center}
		\caption{Cost comparison with standard system.}\label{tab:CostComp}
		\begin{tabular}{ccc}
			\hline
			Unit Quantities (1000 units) & Second-life system & Standard system\\
			\hline\hline
			Li-ion cells & \$ 0.00 & \$ 8.00\\
			Protection circuit & \$ 0.00 & \$ 5.56 \\
			PCB & \$ 1.28 & \$ 1.28 \\
			Electronic components & \$ 20.67 & \$ 9.70 \\
			Assembly & \$ 8.56 & \$ 8.56 \\
			Enclosure & \$ 6.95 & \$ 6.95 \\
			\hline
			Total system cost & \$ 37.47 & \$ 40.05 \\
			Net present value (lifetime: 10 years) & \$ 37.47 & \$ 61.39 \\
			\hline
		\end{tabular}
	\end{center}
\end{table}

This initial financial analysis reveals how the savings due to second-life Li-ion cells can offset the additional costs incurred by the above described system. The following cost estimates are based on 1000 unit quantities and given in USD \$. The cost comparison of a four cell second-life system with a standard four cell Li-ion energy storage system are listed in Table \ref{tab:CostComp}. The largest cost savings of the proposed system result from the Li-ion cells. The cost of new cells for a standard system were estimated at \$ 2.00 per cell, according to an average cost obtained from online retailers such as www.alibaba.com and www.amazon.com. The largest cost factor of the proposed system are the electronic components. However, since the BMS is integrated in the second-life system, there are no additional costs for protection circuits, which are necessary to ensure safe operation in a standard Li-ion system. The cost of assembly and enclosure are equal for the second-life and the standard system, since we assume equal device dimensions. Overall, the total cost of the proposed system is estimated to be \$ 2.59 or 7\% lower than that of an equivalent standard Li-ion based system. Although the difference in fabrication cost between the two devices may appear small, the financial advantages of the second-life system become more apparent when the system lifetime is considered. At a device lifetime of 10 years and a Li-ion cell lifetime of three years in a standard system, the net present value of the proposed system is \$ 23.92 or 64\% lower than that of the standard system. This is because new Li-ion cells must be replaced roughly once in three years at a cost of \$ 8.00 for each replacement. This cost is avoided for the proposed system, under the assumption of a free source of second-life Li-ion cells. A discount rate of 2\% over 10 years was assumed for this calculation. The financial benefits of the proposed system are even greater if a maximum power point tracker is included for optimal utilization of the solar PV panel, which can be incorporated into the software of the proposed system but must be purchased separately for a standard system. This is not considered in the cost comparison in Table \ref{tab:CostComp}.

\subsection{Comparison with commercial systems}

Table \ref{tab:SysComp} provides a comparison of the proposed second-life device with similar commercial systems, currently available on the market. In order to establish an equal baseline for this comparison, we complement the above described system with a 5 W solar PV panel, at an assumed additional cost of \$ 5.00. The energy density of the second-life system is calculated at 50 W$\cdot$h/kg assuming 50\% of the nominal cell capacity and 300 g of weight for the device housing, circuitry and connectors. 

\begin{table}[!h]
	\begin{center}
		\caption{Comparison with commercial systems.}\label{tab:SysComp}
		\begin{tabular}{ccccc}
			\hline
			\multirow{2}{*}{System} & Battery & PV power & Energy density & Cost\\
			& technology & [W] & [Wh/kg] & [USD] \\
			\hline\hline
			Second-life system & Li-ion & 5 & 65 & \$ 42.47 \\
			BBOXX & Lead-acid & 7 & 9 & \$ 80.00 \\
			Panasonic Solar Lantern & NiMH & 3.5 & 50 & \$ 50.00\\
			\hline
		\end{tabular}
	\end{center}
\end{table}

BBOXX is a lead-acid based technology with low energy density (over 5 times the weight of the other two systems) and relatively high cost at \$ 80.00 (88\% higher than the second-life system). The Panasonic Solar Lantern is based on a Nickel-metal hydride (NiMH) battery and features an integrated LED light. The system cost is 18\% higher than that of the second-life system. Neither of the two commercial systems are scalable to higher energy storage capacities. 

This comparison elucidates that the proposed energy storage device based on second-life Li-ion cells is more practical and lower in cost than comparable commercial systems. Based on this analysis we conclude that a profit margin of 15\% on the production cost of the second-life system is possible, while maintaining a competitive financial advantage over comparable commercial systems.

\section{Conclusion}\label{sec:Conclusion}

The power modules and BMS algorithms presented in this contribution enable the construction of an electricity storage device comprised of reused Li-ion cells of various types and capacities. Hardware tests conducted on three interconnected power modules demonstrate the capability of the system to compensate for different cell capacities and voltages, as well as the failure of individual cells. During all tests, the output voltage was successfully maintained at the desired level of 11.3 V with a maximum variance of 34 mV. A BMS algorithm was designed to regulate cell currents based on cell capacities. The algorithm was implemented in a MATLAB Simulink model and used to simulate repeated charge and discharge cycles of three cells with different capacities. The simulation results showed that the cycle times of the three cells converge after two cycles, which proves the viability of the BMS algorithm for an initial synchronization of cells with different capacities. Future work includes refining the BMS algorithm to address the issues of different cell chemistries and degradation during long term operation. The algorithm will then be implemented on the micro-controllers of the power modules and tested on real Li-ion cells of various capacities and chemistries.

The fabrication cost of the proposed device for a 4 cell system was estimated at \$ 37, 7\% lower than the cost of an equivalent device comprised of new Li-ion cells. Accounting for the system cost over a lifetime of 10 years amounts to savings of up to 64\% for the second-life device. Compared to similar commercially available systems, the fabrication cost of the second-life device is between 18\% and 88\% lower than the retail cost the commercial systems and features equal or higher energy densities. Judging from this latter cost comparison, it was concluded that a profit margin of 15\% on the fabrication cost is possible in order to maintain competitive product pricing.

\bibliographystyle{IEEEtran}
\bibliography{references-GHTC}

\end{document}